\documentclass[numbers,preprint,5p]{elsarticle}
\pdfoutput=1 

\usepackage{amssymb}

\usepackage{amsmath}

\usepackage{nomencl}
\makenomenclature
\usepackage[colorlinks=true,linkcolor=black, citecolor=blue, urlcolor=blue]{hyperref}

\usepackage{draftwatermark}
\SetWatermarkText{accepted version}
\SetWatermarkScale{0.7}

\usepackage{lineno}
\newcommand*\patchAmsMathEnvironmentForLineno[1]{%
  \expandafter\let\csname old#1\expandafter\endcsname\csname #1\endcsname
  \expandafter\let\csname oldend#1\expandafter\endcsname\csname end#1\endcsname
  \renewenvironment{#1}%
     {\linenomath\csname old#1\endcsname}%
     {\csname oldend#1\endcsname\endlinenomath}}%
\newcommand*\patchBothAmsMathEnvironmentsForLineno[1]{%
  \patchAmsMathEnvironmentForLineno{#1}%
  \patchAmsMathEnvironmentForLineno{#1*}}%
\AtBeginDocument{%
\patchBothAmsMathEnvironmentsForLineno{equation}%
\patchBothAmsMathEnvironmentsForLineno{align}%
\patchBothAmsMathEnvironmentsForLineno{flalign}%
\patchBothAmsMathEnvironmentsForLineno{alignat}%
\patchBothAmsMathEnvironmentsForLineno{gather}%
\patchBothAmsMathEnvironmentsForLineno{multline}%
}

\begin{document}

\begin{frontmatter}

\title{Deriving environmental contours from highest density regions}

\author[label1,label2]{Andreas F. Haselsteiner\corref{cor1}}
\address[label1]{University of Bremen, Faculty of Production Engineering, BIK -- 	
Institute for Integrated Product Development, 28359 Bremen, Germany}
\address[label2]{ForWind -- Center for Wind Energy Research of the Universities of Oldenburg, Hannover and Bremen, 26129 Oldenburg, Germany}
\cortext[cor1]{Corresponding author}
\ead{a.haselsteiner@uni-bremen.de}

\author[label1,label2]{Jan-Hendrik Ohlendorf}
\ead{johlendorf@uni-bremen.de}

\author[label3]{Werner Wosniok}
\ead{wwosniok@math.uni-bremen.de}
\address[label3]{University of Bremen, Faculty of Math and Computer Science,
Institute of Statistics, 28359 Bremen, Germany }

\author[label1,label2,label4]{Klaus-Dieter Thoben}
\address[label4]{BIBA –- Bremer Institut f{\"u}r Produktion und Logistik GmbH, 28359 Bremen, Germany}
\ead{tho@biba.uni-bremen.de}

\begin{abstract}
Environmental contours are an established method in probabilistic engineering design, especially in ocean engineering. The contours help engineers to select the environmental states which are appropriate for structural design calculations. Defining an environmental contour means enclosing a region in the variable space which corresponds to a certain return period. However, there are multiple definitions of environmental contours for a given return period as well as different methods to compute a contour. Here, we analyze the established approaches and present a new concept which we call highest density contour (HDC). We define this environmental contour to enclose the highest density region (HDR) of a given probability density. This region occupies the smallest possible volume in the variable space among all regions with the same included probability, which is advantageous for engineering design. We perform the calculations using a numerical grid to discretize the original variable space into a finite number of grid cells. Each cell's probability is estimated and used for numerical integration.  The proposed method can be applied to any number of dimensions, i.e. number of different variables in the joint probability model. To put the highest density contour method in context, we compare it to the established inverse first-order reliability method (IFORM) and show that for common probability distributions the two methods yield similarly shaped contours. In multimodal probability distributions, however, where IFORM leads to contours which are difficult to interpret, the presented method still generates clearly defined contours.
\end{abstract}


\end{frontmatter}


\printnomenclature
\nomenclature{HDC}{Highest density contour [-]}
\nomenclature{HDR}{Highest density region [-]}
\nomenclature{IFORM}{Inverse first order reliability method [-]}
\nomenclature{$H_s$}{Significant wave height, random variable [m]}
\nomenclature{$h_s$}{Significant wave height, realization [m]}
\nomenclature{$H_{s\oint}$}{Maximum significant wave height along the contour [m]}
\nomenclature{$H_{s,25}$}{25-year return value of the significant wave height based on its marginal distribution [m]}
\nomenclature{$T_z$}{Zero-upcrossing period, random variable [s]}
\nomenclature{$t_z$}{Zero-upcrossing period, realization [s]}
\nomenclature{$T_{z\oint}$}{Maximum zero-upcrossing period along the contour [s]}
\nomenclature{$\alpha$}{Exceedance probability [-]}
\nomenclature{$Pr()$}{Probability function [-]}
\nomenclature{$X$}{Random variable in original space [-]}
\nomenclature{$x$}{Realization of the random variable in original space [-]}
\nomenclature{$M$}{Random variable in general variable space [-]}
\nomenclature{$m$}{Realization of the random variable in general variable space [-]}
\nomenclature{$p$}{Number of variables / dimensions [-]}
\nomenclature{$f()$}{Probability density function [-]}
\nomenclature{$j$}{Dimension index [-]}
\nomenclature{$f_m$}{Minimum probability density of the enclosed region / constant probability density along the contour [-]}
\nomenclature{$f_m^*$}{Normalized minimum probability density [-]}
\nomenclature{$T$}{Return period [years]}
\nomenclature{$T_p$}{Spectral peak period [s]}
\nomenclature{$U$}{Random variable in standard normal space [-]}
\nomenclature{$u$}{Realization of the random variable in standard normal space  [-]}
\nomenclature{$\beta$}{Radius in $U$-space used in IFORM [-]}
\nomenclature{$n$}{Total number of environmental states in a given time period [-]}
\nomenclature{$r_{M0}$}{Reference point [-]}
\nomenclature{$\theta$}{Angle [$\deg$]}
\nomenclature{$\alpha_W$, $\beta_W$, $\gamma_W$}{Parameters of a Weibull distribution [-]}
\nomenclature{$N()$}{Normal distribution [-]}
\nomenclature{$\mu_2$, $\sigma_2$}{Parameters of a normal distribution [-]}
\nomenclature{$LN()$}{Log-normal distribution [-]}
\nomenclature{$\widetilde{\mu}_{Hs}$, $\widetilde{\sigma}_{Hs}$}{Parameters of a log-normal distribution [-]}
\nomenclature{$a_1$, $a_2$, $a_3$, $b_1$, $b_2$, $b_3$}{Fitted parameters of the conditional distribution [-]}
\nomenclature{$C$}{Set making up the environmental contour [-]}
\nomenclature{$R$}{Set enclosed by the environmental contour (highest density region) [-]}
\nomenclature{$\mathcal{F}$}{Failure region [-]}
\nomenclature{$k$, $l$}{Grid cell index [-]}
\nomenclature{$K$, $L$}{Number of grid cells in the respective dimension  [-]}
\nomenclature{$F()$}{Cumulative distribution function [-]}
\nomenclature{$\bar{F}(f_m)$}{Probability enclosed by a contour of $f_m$ probability density [-]}
\nomenclature{$\bar{f}_{X1}$}{Cell-averaged probability density in dimension 1 [-]}
\nomenclature{$\bar{f}_{X2\vert X1}$}{Cell-averaged probability density in dimension 2  conditional on $x_1$ [-]}
\nomenclature{$\bar{f}$}{Cell-averaged joint probability density [-]}
\nomenclature{$P_f$}{Failure probability [-]}
\nomenclature{$p_{Hs}$}{Mixture coefficient [-]}
\nomenclature{$z$}{Number of components [-]}


\section{Introduction}
\subsection{Purpose of environmental contours}
\label{subsec1}
Engineers have to design any marine structure in such a way that it is able to withstand the loads induced by the environment. As the environment, i.e. wind, waves and currents, continually change and cannot be predicted for long periods of time, the environment is often modeled stochastically by defining probability density functions, $f(x_j)$. Then, the structure is designed to withstand all but some extremely rare environmental states, e.g. all waves with significant wave heights, $H_s$, less than a threshold, $h_s$, with a cumulative probability or \textit{exceedance probability} of $\alpha$, i.e. $Pr(H_s \leq h_s)=1-\alpha$ or $Pr(H_s > h_s)=\alpha$. In general notation for any random variable, $X_1$, there exists a threshold, $x_1$, which fulfills 
\begin{equation}
    F(x_1) = Pr(X_1 \leq x_1)=\int_{-\infty}^{x_1} f(x) dx = 1-\alpha.
\end{equation}
The exceedance probability, $\alpha$, corresponds to a certain \textit{recurrence} or \textit{return period}, $T$, which describes the average time period between two consecutive environmental states above the threshold, $x_1$. The threshold is called \textit{return value}. For example, to comply with standards a marine structure such as an offshore wind turbine is required to withstand significant wave heights, $H_s$, with a return period, $T$, of 50 years \citep{iec:2009}.
\par
\begin{figure}
    \centering
    \includegraphics{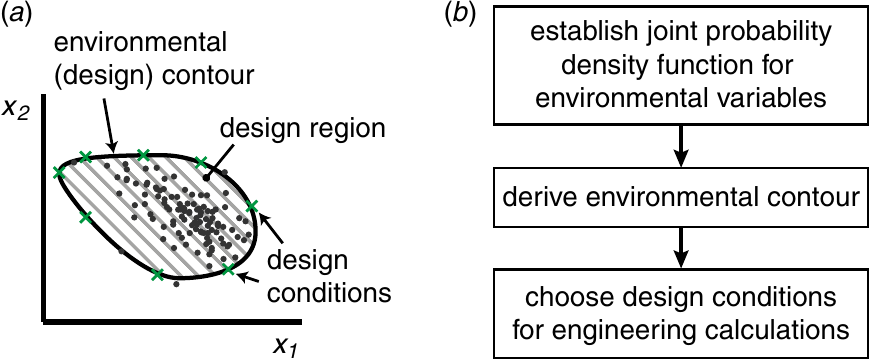}
    \caption{\textbf{Concept of an environmental contour.} (a) The environmental contour encloses all variable combinations which must be considered in the design process (the design region). (b) Flowchart describing the design process utilizing an environmental contour.}
    \label{fig:environmental_contour_general_concept}
\end{figure}

Often, however, structural safety depends not only on one variable, but on the occurrence of combinations of $p$ variables, $\{X_j\}_{j=1}^p$. When two variables are of importance, e.g. significant wave height, $H_s$, and spectral peak period, $T_p$, a joint probability density function can be defined and an environmental contour can be calculated which encloses the subset (or \textit{region}) of environmental states that the structure has to be designed for. Here, we call this region \textit{design region} (Fig. \ref{fig:environmental_contour_general_concept}). Often the most critical structural response is associated with very high or low values of environmental variables, i.e. with environmental conditions located at the boundary of the design region. Consequently, standards allow engineers to calculate structural responses for a limited set of environmental \textit{design conditions} along the contour instead of requiring engineering calculations based on a high number of possible variable combinations spread over the complete design region \citep{dnv:2010environment}. If there are more than two variables the concept of environmental contours leads to environmental surfaces (3 variables) or environmental manifolds ($>3$ variables). Here, for simplicity we also refer to these as environmental contours. 
\subsection{Different definitions and methods}
\begin{figure}
    \centering
    \includegraphics{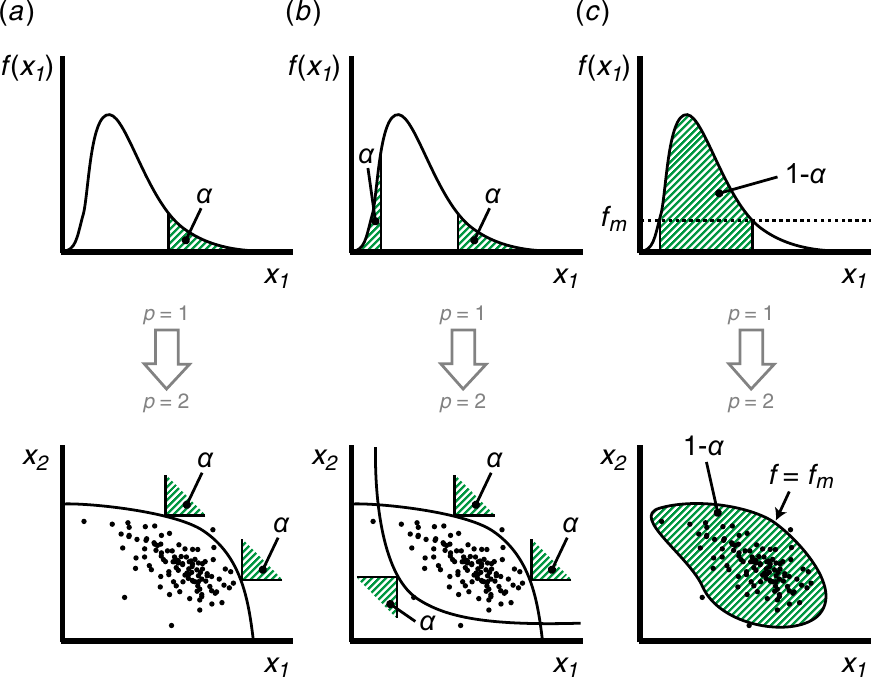}
    \caption{\textbf{Different definitions of environmental contours and their basis in univariate probability distributions.} (a--c) Top: Univariate probability distributions ($p=1$). Bottom: Example data and contours based on two-dimensional joint probability distributions ($p=2$). (a) One-sided exceedance. (b) Two-sided exceedance. (c) Highest density regions (HDRs) with a minimum probability density, $f_m$.}
    \label{fig:different_exceedance_principles}
\end{figure}

\begin{figure}
    \centering
    \includegraphics{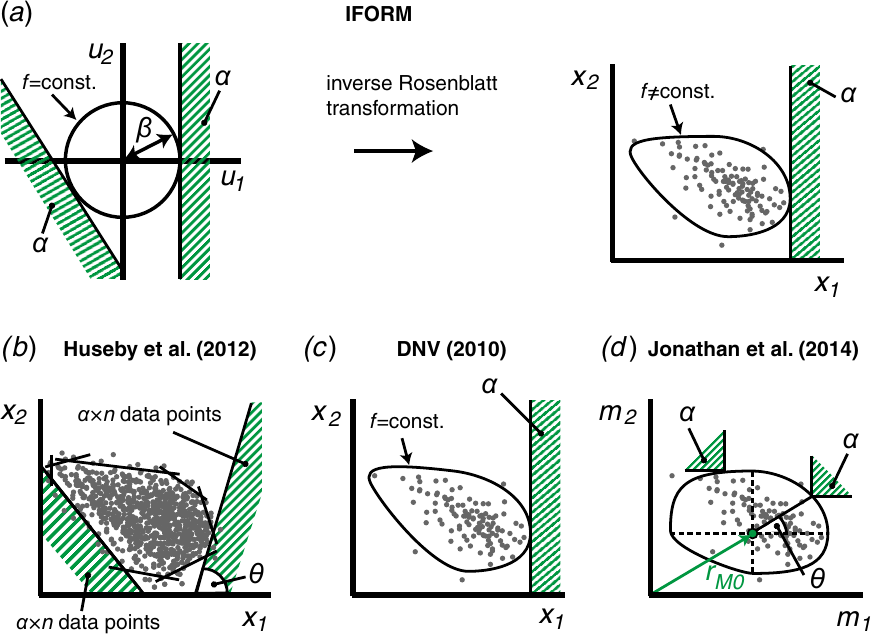}
    \caption{\textbf{Established methods to calculate environmental contours.} (a) \textit{Inverse first-order reliability method} (IFORM) \citep{winterstein:1993}. The contour is defined as a circle in standard normal variables, $U_j$. The points along the circle have to be transformed to the original environmental variables, $X_j$. (b) \citet{huseby:2013} \textit{Monte Carlo contour}. The contour is computed with the environmental variables, $X_j$, directly. (c) Constant probability density contour described by \citet{dnv:2010environment}. By its definition it is unclear how much probability is enclosed by the contour. (d) \citet{jonathan:2014} constant exceedance probability contour. The calculation can be done in a general set of variables, $M_j$, e.g. in the $X$- or $U$-space. In comparison to the other methods a different definition for the exceedance region is used (compare shaded areas).}
    \label{fig:different_methods_calculation}
\end{figure}

As there are different mathematical definitions for environmental contours one has to further specify which kind of environmental contour is being calculated. Different concepts of environmental contours lead to different design loads and consequently to different structural responses \citep{armstrong:2015}. Originally, environmental contours arose from the concept of return values in univariate probability density functions which are calculated based on one-sided exceedance over threshold (Fig. \ref{fig:different_exceedance_principles}\textit{a}). Consequently, a logical definition for an environmental contour is (i) constant one-sided exceedance in all directions of the \textit{p}-dimensional variable space, $Pr(X_1>x_1, X_2>x_2, ..., X_p>x_p)=\alpha$. The bottom panel in Fig. \ref{fig:different_exceedance_principles}\textit{a} shows the contour for the two-dimensional joint distribution of $X_1$ and $X_2$. However, for design purposes not only the highest values of a variable can be of interest, but also the lowest. For example, when designing an offshore structure, low values of the peak period, $T_p$, have to be considered as the structure's natural frequencies can be either higher or lower than the average peak period. Consequently, another possible definition for an environmental contour is (ii) two-sided exceedance over threshold (Fig. \ref{fig:different_exceedance_principles}\textit{b}; e.g. \citep{jonathan:2014}). A third possibility is to define an environmental contour to have (iii) constant probability density, $f_m$, along its path enclosing the most likely environmental states (Fig. \ref{fig:different_exceedance_principles}\textit{c}). In this case a $T$-year return period means that on average every $T$ years an environmental state with a probability density less than $f_m$ occurs. In the broader statistics literature the variable region enclosed by such a contour is called a \textit{highest density region} (HDR) \citep{hyndman:1996}. Although HDRs are a logical concept for environmental contours, yet no author has strictly followed this definition. The \textit{design curve} introduced by \citet{haver:1987} is a related concept since it is a line of constant probability density, but only one-sided exceedance is considered. The constant probability density approach described by \citet{dnv:2010environment} does define a fully closed contour of constant probability density. However, it is designed in such a way that it is unclear how much probability is enclosed by the contour. Instead the contour's probability density, $f$, is chosen to be the joint probability density of the $(x_1,x_2)$-variable combination with $x_1$ = return value based on the marginal $x_1$-distribution and $x_2$ = an associated $x_2$ value (Fig. \ref{fig:different_methods_calculation}c). \citet{leira:2008}, however, has indeed used the HDR definition but only after a transformation of the original variables into standard normal space. When transforming the contour back to the original variable space the constant probability density is not preserved. Here we will compute contours strictly following the HDR definition.
\par
Besides these different definitions of types of environmental contours there exist different methods to calculate a given type of environmental contour. The traditional and probably most used approach is the so-called \textit{inverse first-order reliability method} (IFORM) \citep{winterstein:1993,haver:2009}. It is a standard design practice for a wide range of marine engineering applications where extreme sea states are of interest \citep{dnv:2010environment}. These are for example ships \citep{dnvgl:2015ships}, offshore wind turbines \citep{iec:2009}, floating structures \citep{dnv:2010floating} or wave energy converters \citep{dnv:2008waveenergy,dnv2014:steelstructures}. Using IFORM one defines the contour in standard normal space, $U_j$, instead of the original environmental variable space, $X_j$. Thus, one first defines a circle with a radius, $\beta$, in the $U$-space (Fig. \ref{fig:different_methods_calculation}\textit{a}). The radius corresponds to the return period and increases with longer periods. Then one transforms the points along the circle to the original variable space leading to the environmental contour.This transformation is done via the inverse \textit{Rosenblatt transformation} \citep{rosenblatt:1952}.  As its name implies IFORM is a reliability method and is based on the idea that the \textit{exceedance region} approximates the \textit{failure region}, $\mathcal{F}$, of a structure (and the exceedance probability, $\alpha$, approximates the structure's \textit{failure probability}, $P_f$; see \citep{madsen:1986}). Contours based on IFORM are widely used and have been published e.g. by \citet{saranyasoontorn:2004,leira:2008,baarholm:2010,li:2015,myers:2015,valamanesh:2015,eckert-gallup:2016}.
\par
\citet{huseby:2013}, however, pointed out that the Rosenblatt transformation introduces errors as failure probabilities, $P_f$, can be underestimated or overestimated on a case by case basis. Therefore, they introduced an alternative method to calculate environmental contours in the original variable space. Following their method, one first carries out a Monte Carlo simulation to generate a high number of sea states based on a given joint probability distribution model. Then one chooses an angle, $\theta$, defining a line (in two dimensions, $p=2$) and varies its position such that it divides the variable space into one halfspace containing the majority of data points and the other halfspace containing the data points representing the exceedance probability, $\alpha \times n$ (with $n$ being the total number of simulated environmental states, Fig. \ref{fig:different_methods_calculation}\textit{b}).
By iterating this procedure over a finite number of angles, $\theta \in [0, 360)$, the resulting lines can be connected to an environmental contour. This new approach has been picked up in several recent publications, e.g. to compare the approach to the traditional IFORM method \citep{vanem:2015}, to compare different statistical models  \citep{vanem:2016} or to decrease the required process time \citep{huseby:2014}. While the Monte Carlo method overcomes the problems caused by the Rosenblatt transformation it requires the simulation of environmental states which is computationally more expensive than the simple IFORM calculations. Further, by its definition the method cannot generate concave contours. 
\par
\citet{jonathan:2014} define and calculate environmental contours yet differently. Using clear mathematical notation they find a contour with constant exceedance probability, $Pr(X_1>x_1, X_2>x_2)=\alpha$ (notation for two dimensions, $p=2$). Thus, instead of finding halfspaces which are tangential to the contour, their exceedance regions have finite boundaries for each variable leading to outwards radiating rectangles in a two-dimensional Cartesian coordinate system (Fig. \ref{fig:different_methods_calculation}\textit{d}). Consequently, in contrast to IFORM and the Monte Carlo approach the method does not try to match the exceedance region with the failure region and thus separates the concept of an environmental contour from a structure's failure function. Following this method one first chooses a reference point, $r_{M0}$. Then one defines a line which passes through that point at an angle, $\theta$, to the abscissa. Lastly one  finds the position along the line which satisfies $Pr=\alpha$. Repeating this procedure over a full circle, $\theta \in [0, 360)$, one finds the environmental contour. The method can be applied in any variable space, $M_j$, e.g. in original variables, $X_j$, or standard normal variables, $U_j$. Further, besides fully closed contours, one-sided exceedance is also considered by the authors. One can combine the method with using modern conditional extreme models \citep{heffernan:2004} as demonstrated by \citet{jonathan:2010,jonathan:2014}. The method disconnects the environmental state statistics from any particular structural problem which makes it a more general approach to define a $T$-year set of environmental states for any further use of these data. However, like the reliability methods, it defines multiple $\alpha$-exceedance regions in the variable space of a single probability model. While in reliability methods the idea is that one of these multiple exceedance regions overlaps with the failure region this is not the case with the $Pr(X_1>x_1, X_2>x_2)=\alpha$ definition. Thus, if a contour is defined independently of the concept of failure regions, it seems more meaningful to define $\alpha$ to be the probability of a single region (instead of having multiple regions with $\alpha$ probability content).
\par
Motivated by the individual advantages the described contour calculation methods have, here, we introduce contours enclosing highest probability density regions which we compute using numerical integration. We continue the idea introduced by \citet{jonathan:2014} of decoupling the exceedance region from the structure's failure region, but go one step further and do not define any kind of outwards radiating exceedance region. Instead, we choose to find a contour which encloses the most likely environmental states which together make up a defined probability of $1-\alpha$. The proposed method allows us to define the contour in the original variable space and can be used for any number $p$ of dimensions. By discretizing the variable space into a finite number of grid cells and using numerical integration techniques any probability distribution can be evaluated, e.g common parametric sea state joint probability distributions \citep{vanem:2012}, nonparametric models \citep{eckert-gallupandmartin:2016} or extreme value models which can have discontinuities at the threshold \citep{scarrott:2012}. 
 Similar as being done e.g. in computational fluid dynamics \citep{celik:2008} we demonstrate that with a sufficiently small grid cell size the solution is grid independent. 
\section{Data}
In order to compare our environmental contour approach to similar methods we use the 3-hour sea state model presented by \citet{vanem:2012}. They use a fitted joint model for significant wave height, $H_s$, and zero-upcrossing period, $T_z$. Based on their model environmental contours have been calculated using both the traditional IFORM method \citep{vanem:2012} and the newer Monte Carlo method \citep{huseby:2013}. The joint model was derived from one particular location in the ERA-Interim data set \citep{dee:2011}. Significant wave height, $H_s$, is modeled as a 3-parameter Weibull distribution with the parameters $\alpha_W$ (scale), $\beta_W$ (shape) and $\gamma_W$ (location):
\begin{equation}
\begin{split}
    &f_{Hs}(h_s) =\\ &\dfrac{\beta_W}{\alpha_W}\left( \dfrac{h_s-\gamma_W}{\alpha_W}\right)^{\beta_W -1}
    \exp \left[-\left( \dfrac{h_s-\gamma_W}{\alpha_W} \right)^{\beta_W} \right]; \enspace h_s \geq \gamma_W .
\end{split}
\end{equation}
Based on a least squares fit the parameters are $\alpha_W=2.776$, $\beta_W=1.471$ and $\gamma_W=0.8888$ \citep{vanem:2012}.
\par
The zero-upcrossing period, $T_z$, is modeled to follow a log-normal distribution, $LN$:
\begin{equation}
\begin{split}
    &f_{T_z|H_s}(t_z|h_s) =\\
    &LN(\widetilde{\mu}_{Hs}, \widetilde{\sigma}_{Hs}^2) =
    \dfrac{1}{t_z\widetilde{\sigma}_{Hs} \sqrt{2\pi}}\exp \left[ \dfrac{-(\ln t_z - \widetilde{\mu}_{Hs})^2}{2\widetilde{\sigma}_{Hs}^2}\right] .
\end{split}
\label{eq:log-normal}
\end{equation}
The distribution's parameters, $\widetilde{\mu}_{Hs}$ and $\widetilde{\sigma}_{Hs}$, are conditional on the significant wave height, $H_s$, and are modeled as 3-parameter functions: 
\begin{align}
    \widetilde{\mu}_{Hs} (h_s) &= a_1 + a_2 h_s^{a_3} ,\\
    \widetilde{\sigma}_{Hs} (h_s) &= b_1 + b_2 \exp (b_3 h_s) .
\label{eq:conditional_log-normal}
\end{align}
In this case they are estimated to be $a_1=0.1000$, $a_2=1.489$, $a_3=0.1901$, $b_1=0.0400$, $b_2=0.1748$, $b_3=-0.2243$ \citep{vanem:2012}.
\par
Multiplying the marginal distribution of the significant wave height, $H_s$, and the conditional distribution of the zero-upcrossing period, $T_z$, we can calculate the joint distribution:
\begin{equation}
        f_{H_s, T_z}(h_s,t_z) = f_{H_s}(h_s)f_{T_z|H_s}(t_z|h_s) .
\end{equation}
Since the data represent 3-hour sea states, exeedance probability, $\alpha$, for a  $T$-year return period is calculated as
\begin{equation}
\alpha = \dfrac{1}{n} = \dfrac{1}{T \times 365.25 \times 24/3}.
\end{equation}

\section{Highest density contour (HDC)}
\subsection{Analytical definition}
Our goal is to find a contour, $C$, of constant probability density, $f_m$, which encloses a probability of $1-\alpha$, i.e.:
\begin{equation}
\begin{split}
    &C(f_m) = \{x:x \in \mathbb{R}^p, f(x) = f_m\},\\
    &R(f_m) = \{f(x) \geq f_m\}, \\
    &\int_{R(f_m)} f(x) dx = 1 - \alpha.
\end{split}
\label{eq:hdc}
\end{equation}
This contour, $C$, encloses the highest density region, $R$. Therefore we call $C$ \textit{highest density contour} (HDC). A highest density region fulfills two main properties: (i) the probability density of every point inside is at least as large as the probability density of any point outside and (ii) for a given probability content the region occupies the smallest possible volume in the variable space \citep{box:1992}. There is no general analytic solution to find the HDR or HDC, i.e. solving for $C$ or $R$ in Eq. \ref{eq:hdc}. 
\par 
HDRs, however, can be computed based on numerical integration approaches \citep{wright:1986} or Monte Carlo techniques \citep{hyndman:1996}. Environmental contours involve very low $\alpha$ values and are usually based on low-dimensional probability models. Thus, we choose numerical integration over Monte Carlo simulation to compute the highest density contour, $C$. However, if a probability model, which incorporates many environmental variables (high $p$ value), is evaluated numerical integration might become infeasible and Monte Carlo approaches should be used. Here, we use numerical integration and start by discretizing the probability density space into a finite number of equally sized grid cells. In the next section we will evaluate the two-dimensional case, but in the appendix the equations for $p$ dimensions are given.

\subsection{Numerical integration approach in two dimensions}
\begin{figure}
    \centering
    \includegraphics{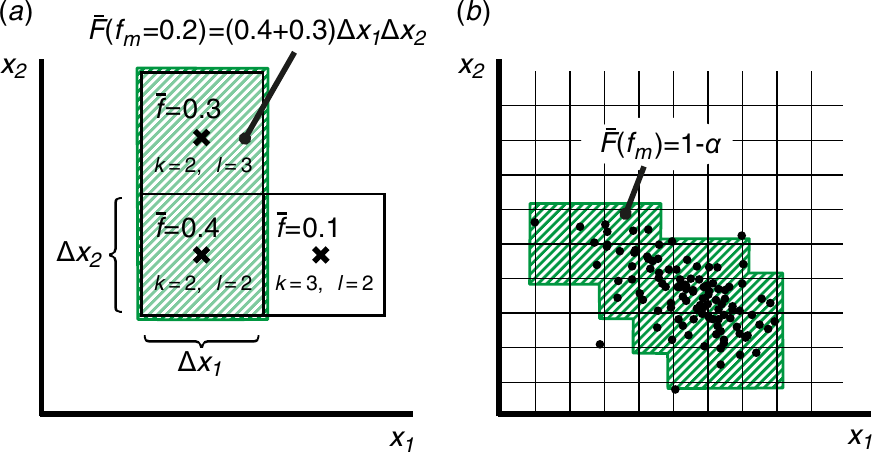}
    \caption{\textbf{Computation of the highest density contour (HDC) using a numerical grid.} Shaded area = HDR, outline = HDC. (a) The variable space is discretized in equally sized grid cells and the average probability density, $\bar{f}$, is calculated for each cell. The probability enclosed by a HDC of $f_m$ probability density is calculated by first finding all cells whose probability density, $\bar{f}$, is greater than or equal the minimum probability density, $f_m$, and then summing up the individual probabilities of these cells. (b) An environmental contour is computed by iteratively finding the minimum probability density, $f_m$, that satisfies $\bar{F}(f_m)=1-\alpha$.}
    \label{fig:concept_numerical_grid}
\end{figure}

The two-dimensional probability space is discretized in $K\times L$ grid cells with a constant size of $\Delta x_1 \times \Delta x_2$ (Fig. \ref{fig:concept_numerical_grid}). Each grid cell's center point, $(x_{1k}, x_{2l})$, is used as the reference position of the cell. Then, based on the cumulative distribution function, $F_{X1}$, the cell-averaged probability density in the first dimension, $\bar{f}_{X1}$, is calculated using central difference:
\begin{equation}
       \bar{f}_{X1}(x_1)=\dfrac{F_{X1}(x_{1} + 0.5 \Delta x_1)-F_{X1}(x_1 -0.5 \Delta x_1)}{\Delta x_1}.
\end{equation}
The cell-averaged probability density in the second dimension, $\bar{f}_{X2\vert X1}$, is calculated similarly:
\begin{equation}
\begin{split}
       &\bar{f}_{X2|X1}(x_2|x_1) =\\
       &\dfrac{F_{X2|X1}(x_2 + 0.5 \Delta x_2)-F_{X_2|X_1}(x_2 - 0.5 \Delta x_2)}{\Delta x_2}.
\end{split}
\end{equation}
While $\bar{f}_{X1}$ is the true cell-averaged probability density in the first dimension, in the second dimension, $\bar{f}_{X2}$ is approximated since the dependence of $F_{X2|X1}$ upon $x_1$ within the grid cell is not accounted for. Instead we fix $x_1$ to the value at the grid cell center, $x_1=x_{1l}$, and therefore assume $F_{X2|X1}$ to be constant from $x_{1l}-0.5\Delta x_1$ to $x_{1l}+0.5\Delta x_1$.
\par 
Multiplying the two individual probability densities yields the cell-averaged joint probability density, $\bar{f}$:
\begin{equation}
       \bar{f}(x_1, x_2)=\bar{f}_{X1}(x_1) \bar{f}_{X2|X1}(x_2|x_1).
\end{equation}
Now we can compute the probability that an event with a minimum probability density of $f_m$ occurs, i.e. we calculate the probability content enclosed by a HDC of $f_m$ probability density. This probability, $\bar{F}(f_m)$, is calculated by summing up the probabilities of all cells which have a probability density greater than or equal $f_m$ (Fig. \ref{fig:concept_numerical_grid}):
\begin{equation}
\begin{split}
    &\bar{F}(f_m)=\\
    &\sum_{k=1}^K \sum_{l=1}^L
    \left\{
        \begin{array}{ll}
          \bar{f}(x_{1k}, x_{2l}) \Delta x_1 \Delta x_2 &\quad \textnormal{if} \enspace \bar{f}(x_{1k}, x_{2l})\geq f_m\\
          0 &\quad \textnormal{if} \enspace \bar{f}(x_{1k}, x_{2l}) < f_m .
        \end{array}
    \right.
\end{split}
\end{equation}
If the joint probability density function is unimodal the grid cells which fulfill $\bar{f} \geq f_m$ make up a single contiguous area. The boundary of this area is a contour which encloses a probability of $\bar{F}$. Using the function $\bar{F}(f_m)$ we can consequently find a contour with a given exeedance probability, $\alpha$, of interest by finding the corresponding minimum probability density, $f_m$:
\begin{equation}
\bar{F}(f_m)=1-\alpha.
\end{equation}
Solving this equation is a root finding problem of a monotonically decreasing function ($\bar{F}(f_m)-1+\alpha=0)$. We solve the equation using Matlab's (version R2015b, The MathWorks, USA) \textit{fzero} function which iteratively finds the root of a nonlinear function. All grid cells fulfilling $\bar{f} \geq f_m$ then approximate the HDR, $R(f_m)$, and the grid cells making up the boundary of the HDR approximate the HDC, $C(f_m)$.
\section{Results and discussion}
\subsection{Properties of the highest density contours}
\begin{figure}
    \centering
    \includegraphics{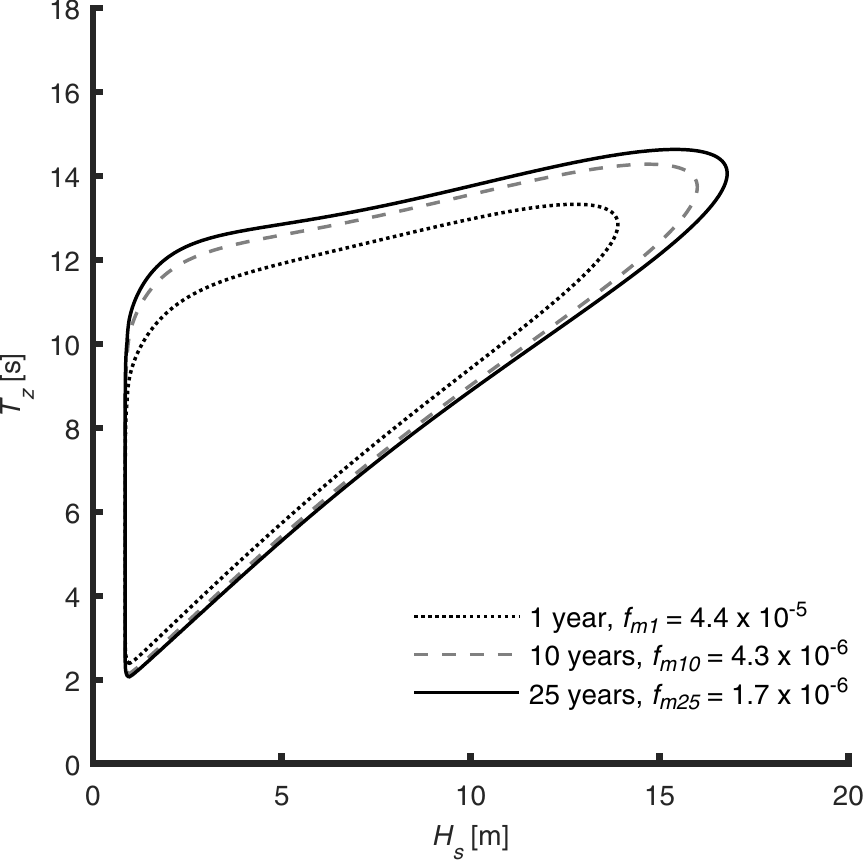}
    \caption{\textbf{Computed highest density contours.} Along the contour probability density, $f_m$, is constant and the enclosed region has a probability of $1-\alpha$ with $\alpha$ corresponding to a given $T$-year return period ($T=1$, 10 or 25 years). Grid cell size is 0.05\,m$\times$0.05\,s. }
    \label{fig:numerical_grid_method_vanem2012}
\end{figure}

\begin{figure*}
    \centering
    \includegraphics{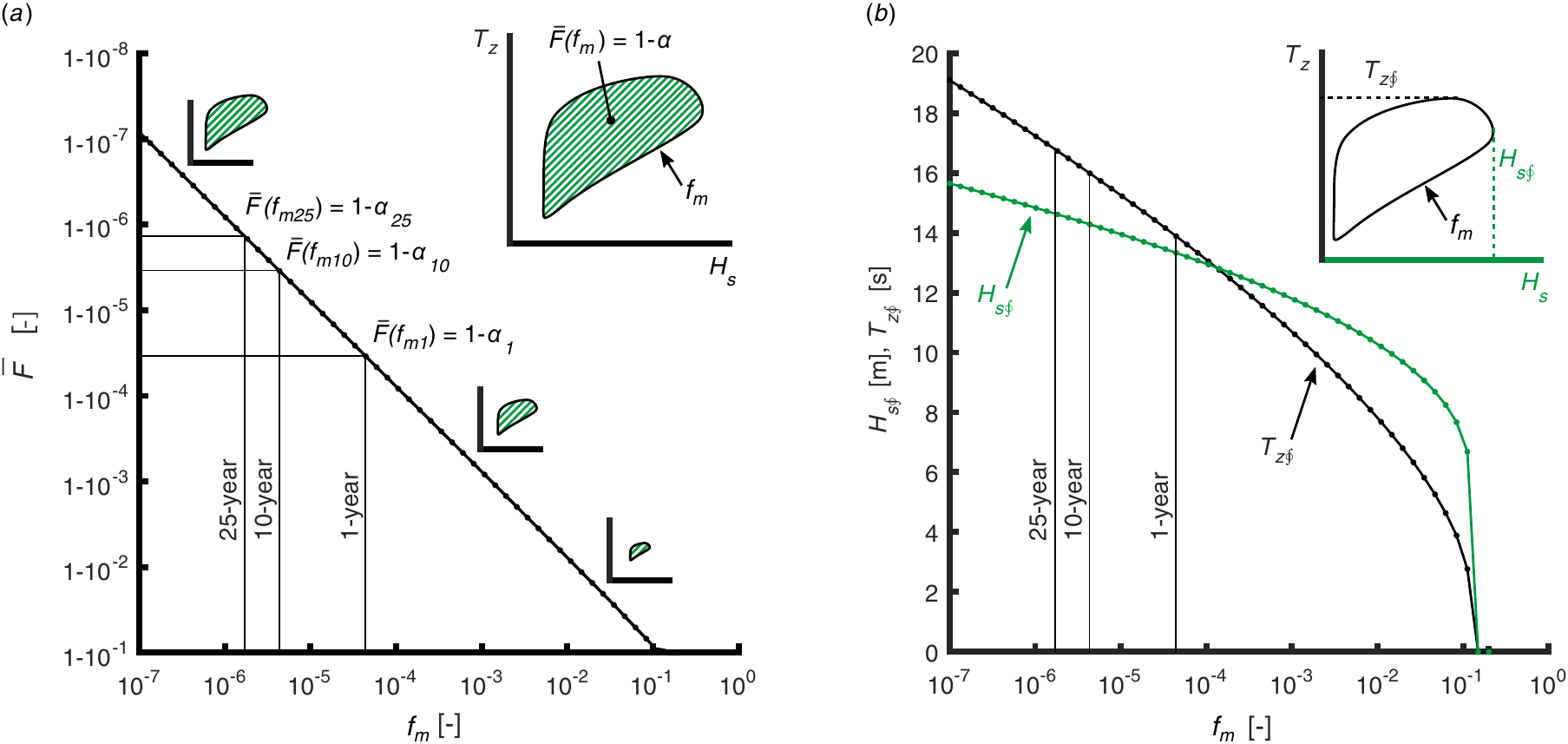}
    \caption{\textbf{Expansion of the highest density contour.} (a) The probability enclosed by the contour, $\bar{F}(f_m)$, is 1 at a minimum probability density of $f_m=0$ and monotonically decreases to $\bar{F}(f_m\approx0.12)=0$. Probabilities corresponding to the 1-,10- and 25-year contour are shown. The inlet illustrates the definition of $\bar{F}$ and $f_m$. (b) Maximum variable values along the contour, $H_{s\oint}$ and $T_{z\oint}$, as a function of minimum probability density, $f_m$. The inlet illustrates that there is no ($H_{s\oint}$,$T_{z\oint}$)-sea state along the contour. Instead, the ($H_{s\oint},T_z)$-sea state has a $T_z$ value different from $T_{z\oint}$ and vice versa.}
    \label{fig:numerical_grid_maxhstz}
\end{figure*}

As done in previous work based on the described joint probability model \citep{huseby:2013,vanem:2012} we compute the 1-, 10- and 25-year environmental contours (Fig. \ref{fig:numerical_grid_method_vanem2012}). The corresponding exceedance probabilities are $\alpha_1=3.42\times 10^{-4}$, $\alpha_{10}=3.42\times 10^{-5}$ and $\alpha_{25}=1.37\times 10^{-5}$ respectively. The computed HDCs have constant probability densities of $f_{m1}=4.4\times 10^{-5}$ (1-year), $f_{m10}=4.3\times 10^{-6}$ (10-year) and $f_{m25}=1.7\times 10^{-6}$ (25-year). Fig. \ref{fig:numerical_grid_maxhstz}\textit{a} shows how the enclosed probability, $\bar{F}$, monotonically decreases with increasing $f_m$ until it reaches $\bar{F}=0$. Since the probability functions we use here (Weibull and log-normal) are unbounded, $\bar{F}$ asymptotically approaches 1 as $f_m$ approaches 0. Fig. \ref{fig:numerical_grid_maxhstz}\textit{b} presents the maximum $H_s$- and $T_z$-values along a contour of constant $f_m$-probability density ($H_{s\oint}$,  $T_{z\oint}$). Longer return periods, $T$, lead to smaller $f_m$-values and consequently to bigger contours with higher $H_{s\oint}$ and  $T_{z\oint}$ values.
\par
\par
\par
\begin{figure*}
    \centering
    \includegraphics{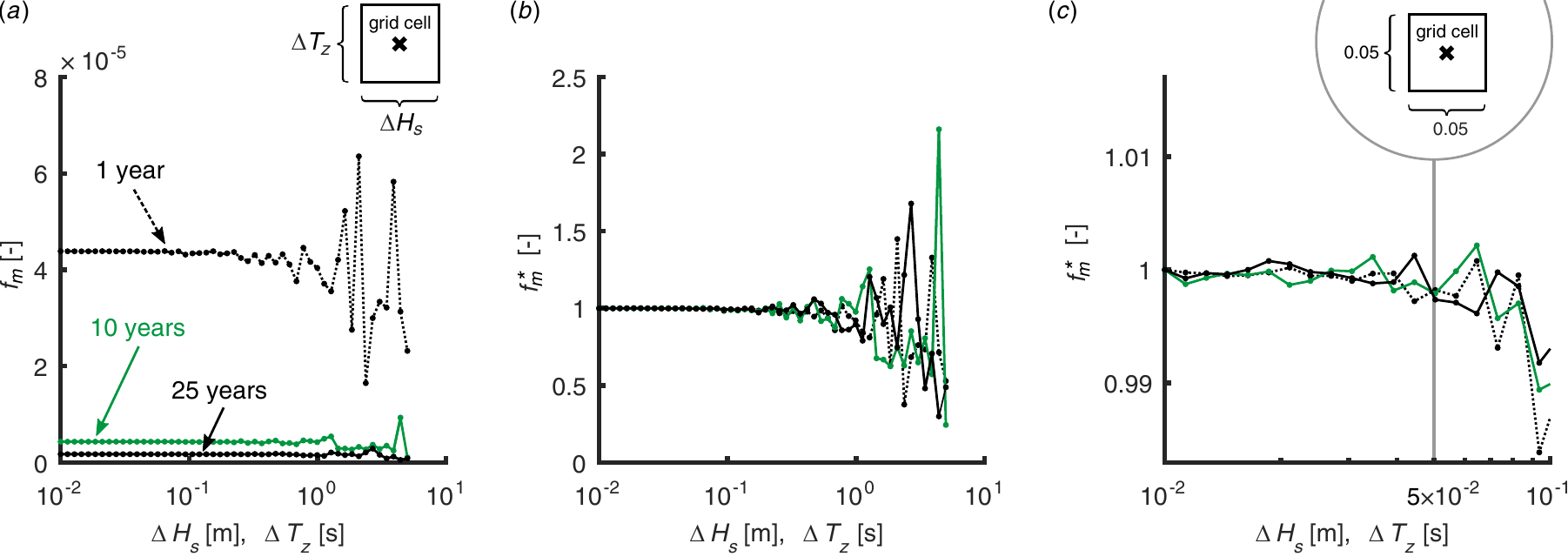}
    \caption{\textbf{Grid independence study.} Quadratic grid cells with sizes ranging from 0.01 to 10 units of grid cell length are tested to evaluate grid convergence. (a) The contour's minimum probability density, $f_m$, for a given return period is sensitive to grid cell size $\Delta H_s$,  $\Delta T_p$. Sensitivity increases with grid cell length. (b) If grid cell size is too big minimum probability density, $f_m$, can be half or double than the converged minimum probability density. Plotted is $f_m^*$ which is minimum probability density, $f_m$, normalized by the converged $f_m$ value. (c) Aiming for grid convergence with an error of less than 1\,\% we use grid cells with dimensions of 0.05\,m$\times$0.05\,s (marked with a vertical line).}
    \label{fig:grid_study}
\end{figure*}

As discretization in general is sensitive to step size we evaluate the contour's robustness with respect to grid cell size $\Delta x_1 = \Delta H_s$, $\Delta x_2 = \Delta T_z$. We analyze how minimum probability density, $f_m$, changes with grid cell size. In all three tested return periods (1-, 10- and 25-year contour) minimum probability density, $f_m$, is roughly constant at small cell sizes and starts to fluctuate with increasing cell size indicating a grid-independent solution can be reached (Fig. \ref{fig:grid_study}\textit{a}). Oversized grid cells can lead to minimum probability density being half or double than the converged minimum probability density (Fig. \ref{fig:grid_study}\textit{b}). For the given probability model we find that convergence is reached at a grid cell size of $H_s=0.05$\,m and $T_z=0.05$\,s. There, deviation to the smallest tested grid cell size is less than 1\,\%, $0.99<f_m^*<1.01$, with $f_m^*$ being  minimum probability density, $f_m$, normalized by the converged $f_m$ value (Fig. \ref{fig:grid_study}\textit{c}).
\subsection{Comparison with IFORM and Monte Carlo contours}
\begin{figure*}
    \centering
    \includegraphics{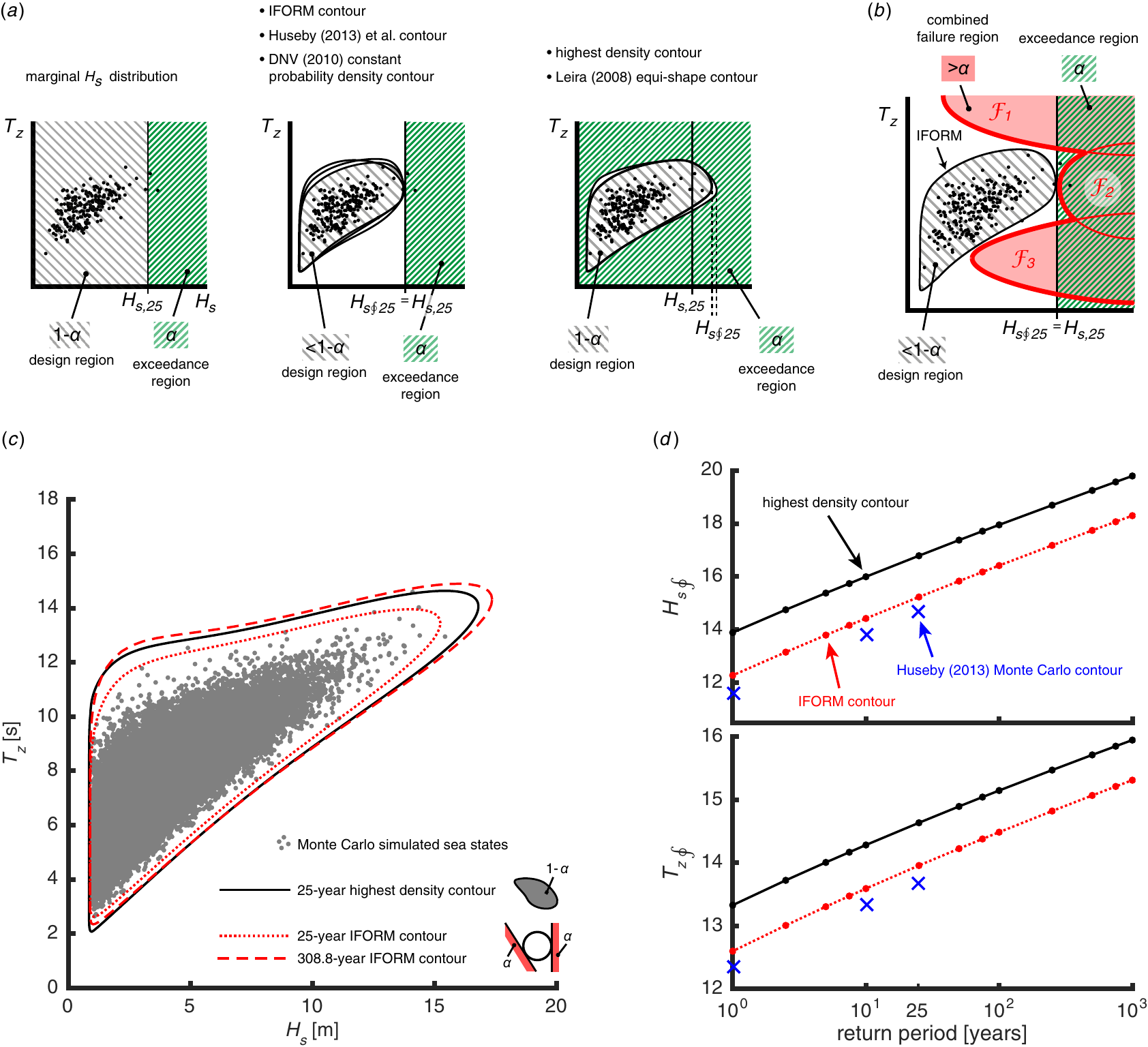}
    \caption{\textbf{Comparison of environmental contours derived using different methods.} (a) Sketches showing expected differences in contour size due to  different definitions. Some contours are defined in such a way that the maximum value along the contour, $H_{s\oint 25}$, is equal the return value of the marginal distribution, $H_{s25}$, (middle). The highest density contour (HDC), however, is defined to enclose $1-\alpha$ probability and thus has a maximum value along the contour which is higher than the return value of the marginal distribution, $H_{s\oint 25} > H_{s25}$, (right). (b) Sketch illustrating an IFORM contour and possible failure regions of a \textit{linear system} of three components, $\mathcal{F}_1$, $\mathcal{F}_2$, $\mathcal{F}_3$. Since the contour contains less than $1-\alpha$ probability the system's failure probability can be greater than $\alpha$. (c) A total of $n=73050$ sea state data points have been Monte Carlo simulated representing a 25-year data set (scatter plot). The 25-year HDC (solid line) and the 25-year IFORM contour (short dashes) have similar shapes, but as expected the HDC is bigger. The 308.8-year IFORM contour or 25-year \textit{equi-shape contour} (long dashes; \citep{leira:2008}) encloses the same amount of probability as the 25-year HDC. (d) Comparison of maximum values along the contour, $H_{s\oint}$ and $T_{z\oint}$. As expected by the different definitions, the HDC has the highest maximum significant wave height, $H_{s\oint}$, and maximum zero-upcrossing period, $T_{z\oint}$.}
    \label{fig:comparison_iform_huseby}
\end{figure*}

For comparison we further compute environmental contours using IFORM based on the same probability model. The highest density contours have similar shapes as the contours calculated with IFORM and the Monte Carlo method (Fig. \ref{fig:comparison_iform_huseby}\textit{c,d}). However, we define a HDC to enclose a probability of $1-\alpha$ while an IFORM contour and a \citet{huseby:2013} Monte Carlo contour each enclose a probability less than $1-\alpha$ since by their definitions multiple regions outside the contour have a probability of $\alpha$  (Fig. \ref{fig:comparison_iform_huseby}\textit{a}). Consequently, the HDC's dimensions in terms of $H_s$ and $T_z$ are bigger in comparison. However, for a fairer comparison we can inflate an IFORM contour and find the $T$-year contour which encloses exactly $1-\alpha$ probability. \citet{leira:2008} showed that this can be done by utilizing the inverted Rayleigh distribution (for two dimensions). The author calls these contours \textit{equi-shape contours}. Here, we find that such a 25-year equi-shape contour corresponds to a 308.8-year IFORM contour. The contour's shape and size is roughly similar to the 25-year HDC. These similarities suggest that the 308.8-year IFORM contour has approximately constant probability density, $f_{m25}$, along the contour.
\par 
To visualize a typical data set, we Monte Carlo simulate 25 years of 3-hour sea states ($n=73050$; gray dots in Fig. \ref{fig:comparison_iform_huseby}\textit{c}). In this particular data set one data point exceeds the HDC while there are multiple data points exceeding the 25-year IFORM contour. The different contour dimensions can also be expressed in terms of maximum $H_s$- and $T_z$-values along the contour ($H_{s\oint}$,  $T_{z\oint }$). While \citet{huseby:2013} report 25-year maxima of $H_{s\oint 25}=14.66$\,m and $T_{z\oint 25}=13.68$\,s for the Monte Carlo contour, here we find $H_{s\oint 25}=16.79$\,m and $T_{z\oint 25} = 14.64$\,s for the HDC and $H_{s\oint 25}=15.23$\,m and $T_{z\oint 25}=13.96$\,s for the IFORM contour (Fig. \ref{fig:comparison_iform_huseby}\textit{d}). Thus, the HDC $H_{s\oint 25}$ value is 10.2~\% higher than the IFORM value and 14.5~\% higher than the Monte Carlo method value. Consequently, from an engineering design point of view the HDC is the most conservative method of the three considered. 
\par 
This does not only apply to the considered distribution, but is a generic property based on the different definitions of these three contours. The IFORM and Monte Carlo contours are defined to contain the return value of the marginal distribution as their highest variable value, i.e. $H_{s\oint 25} = H_{s,25}$ (Fig. \ref{fig:comparison_iform_huseby}\textit{a}). On the other hand, a HDC is defined to enclose $1-\alpha$ probability. Since it does not contain all $H_s$-$T_z$ sea states fulfilling $H_s<H_{s,25}$ (which together would make up $1-\alpha$ probability) it must contain some sea states with $H_s>H_{s,25}$. 
\par 
\par
By the HDC's definition of an enclosed probability of $1-\alpha$, in a random 25-year data set the probability that at least one data point exceeds a 25-year contour is about 63.2\,\%, $1-(1-\alpha_{25})^n\approx0.632$ with $n=25\times 365.25 \times 24/3 = 73050$. Here, exceedance precisely means that this sea state realization is anywhere outside the region enclosed by the contour, $R(f_m)$. Such a sea state occurs on average every 25 years. This simple and clear interpretation is why we have chosen the definition of constant probability density and a probability of $1-\alpha$, i.e. defining the contour to enclose the highest density region. We believe that this definition offers an intuitive and meaningful concept for a $T$-year environmental contour in the engineering design process. If an engineer designs a structure to withstand all sea states inside a $T$-year contour, the structure will be designed for the most likely (extreme) sea states which are expected to occur in $T$ years. Then on average every $T$ years a sea state will occur which the structure is not designed for.
\par 
Alternative concepts with multiple $\alpha$-exceedance regions (see Fig. \ref{fig:different_methods_calculation}\textit{a,b}) are based on the idea of known failure regions in the context of structural reliability methods (see \citep{madsen:1986}). IFORM assumes that a structure's \textit{failure surface} (or \textit{limit state surface}) has a convex shape. It defines the $\alpha$-halfspace exceedance regions in its particular way because in that case the true failure surface can be linearized such that the variable space is separated by a straight line at an angle $\theta$ into a survival region and a failure region (in two dimensions). Then, this failure region overlaps with IFORM's exceedance region. It has the failure probability $P_f=\alpha$ and the survival region the survival probability $1-P_f$. Here, however, we completely separate the idea of describing the environmental conditions from any particular structural problem. Thus, we do not intend to align the $\alpha$-probability exceedance region with a particular failure region.
\par 
 As described IFORM leads to a contour which encloses less than $1-\alpha$ probability and consequently results in less conservative design conditions compared to a HDC. If the structural design, which is developed based on these environmental conditions, has a convex failure surface, the theoretical precondition of IFORM is met. Then in comparison, a HDC can be seen as overly conservative. Thus, if the designer knows that a structure responds with a convex failure surface choosing an IFORM contour is advantageous in the sense that it yields less conservative but still safe design conditions.
 \par 
 While many structures respond with a convex failure surface this precondition for IFORM connects the environmental contour to a certain class of structures. The shape of the failure surface might be unknown beforehand and only becomes apparent during the design process. If it turns out that the failure surface is non-convex and therefore violates IFORM's precondition the designer would need to go one step backwards and define new design conditions by inflating the IFORM contour. By not making use of the properties of possible structural responses the HDC is more conservative, but also more general in its application. It would avoid the need of the described iteration loop in the design process.
 \par 
 Further, a highest density contour is advantageous in the design process of a structural problem of a \textit{system} consisting of multiple \textit{components}.  Consider a \textit{series structure} consisting of $z$ different  components with $z$ different failure functions. In a series structure a failure of one component results in failure of the system \citep{barlow:1975}. Suppose that each component fulfills IFORM's precondition of having a convex failure surface. Nevertheless, the probability contained by the union of all $z$ failure regions, $\mathcal{F}_1 \cup \mathcal{F}_2 ... \cup \mathcal{F}_z$, could exceed $\alpha$ (Fig. \ref{fig:comparison_iform_huseby}\textit{b}). In that case it would be expected that frequenter than every $T$ years an environmental state occurs which leads to failure of some of the components and consequently failure of the system. If an environmental contour containing $1-\alpha$ probability were used to design the components, on the other hand, by definition the system's probability of failure would be less than $\alpha$. Consequently, the system would be expected to survive longer than $T$ years.
 \par 
 A similar example could be given for a single component with multiple failure modes. The three failure regions shown in Fig. \ref{fig:comparison_iform_huseby}\textit{b} would then correspond to different failure modes and the same conclusions as for the series structure could be drawn. These two examples explain why IFORM is primarily aimed at assessing the reliability of one component failing in one particular failure mode. A highest density contour, on the other hand, could be used in these two cases without worrying that any assumptions might be violated.
\subsection{Bimodal mixture model}
Highest density contours can be computed based on any probability distribution. The used definition of constant probability density along the contour, $f_m$, can lead to multiple enclosed subregions for a given return period, $T$, if the probability distribution is multimodal. Here, we demonstrate this by extending the joint $H_s$-$T_z$-probability distribution by a mixture model for the zero-upcrossing period, $T_z$. We use the $H_s$-$T_z$ environmental variables although we are aware that such a $H_s$-$T_z$ distribution might be physically unrealistic. However, for simplicity we build upon the previously described sea state model instead of setting up a new case with a different set of environmental variables. Thus, we keep the log-normal distribution term, $LN(\widetilde{\mu}_{Hs}, \widetilde{\sigma}_{Hs}^2)$, from Eqs. \ref{eq:log-normal}--\ref{eq:conditional_log-normal} and mix it with a normal distribution, $N(\mu_2, \sigma_2^2)$:
\begin{equation}
\begin{split}
    &f_{T_z|H_s}(t_z|h_s) =\\
    &p_{Hs}LN(\widetilde{\mu}_{Hs}, \widetilde{\sigma}_{Hs}^2) + (1-p_{Hs}) N(\mu_2, \sigma_2^2).
\end{split}
\end{equation}
Similar to the parameters $\widetilde{\mu}_{Hs}$ and $\widetilde{\sigma}_{Hs}$ we define the mixture coefficient, $p_{Hs}$, to be conditional on significant wave height, $H_s$. Using an exponential decay function, we let the normal distribution term, $N(\mu_2, \sigma_2^2)$, fade out at high significant wave height, $H_s$:
\begin{equation}
    p_{Hs}(h_s) =  1 - \exp(-3h_s).
\end{equation}
\par
\begin{figure*}
    \centering
    \includegraphics{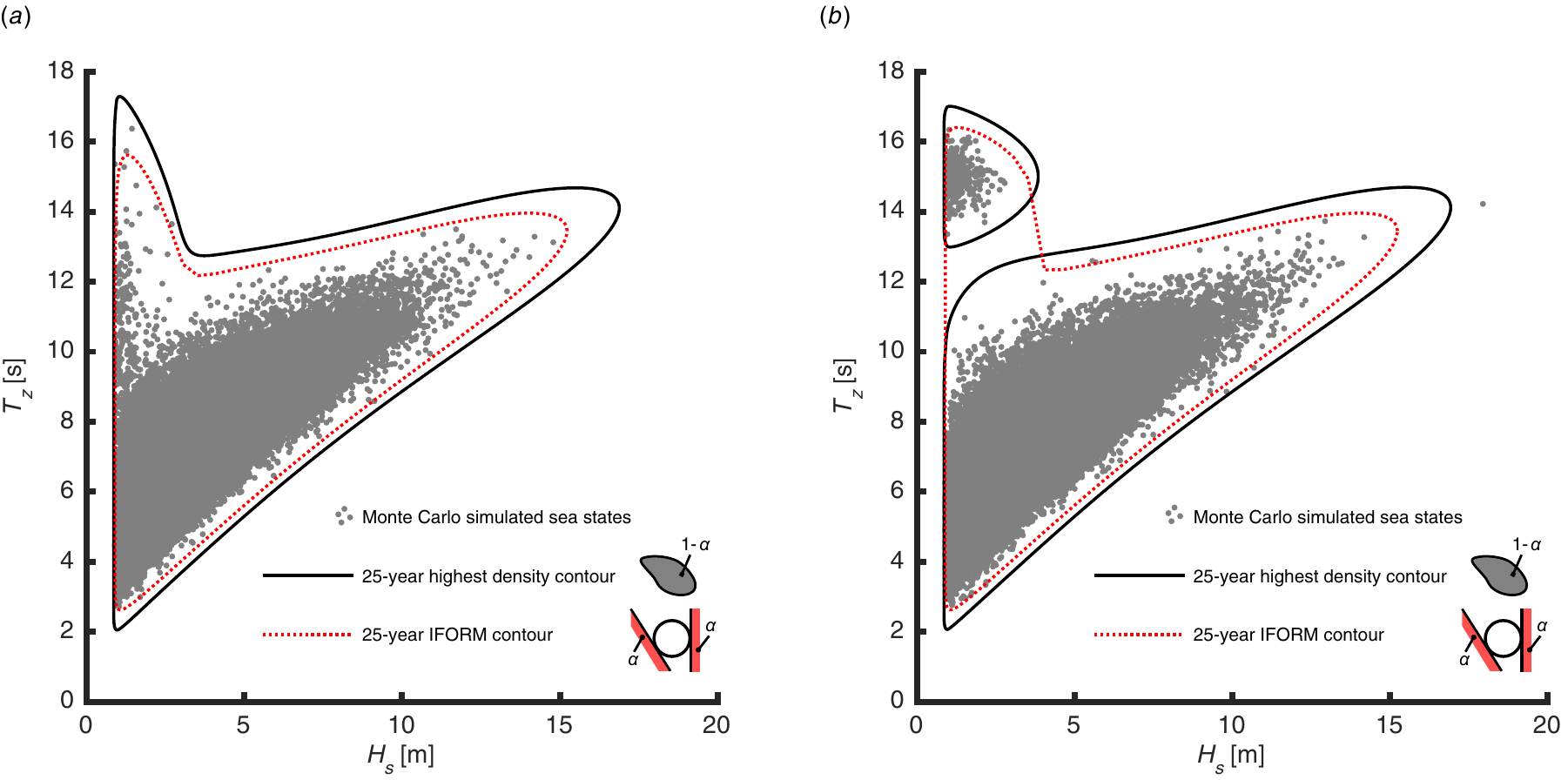}
    \caption{\textbf{Environmental contours for mixture models.} (a) Model~1 has a normal $T_z$-distribution,  $N(\mu_2 = 10$~s, $\sigma_2=2$~s$)$, which smoothly blends into the original $T_z$-log-normal distribution. The highest density contours and IFORM contours have similar shapes. (b) Model~2 has a normal $T_z$-distribution, $N(\mu_2 = 15$~s, $\sigma_2=0.5$~s$)$, which leads to a second probability density maximum. Consequently, the highest density contour encloses two separated subregions. Due to its definition IFORM, however, encloses a single contiguous region.}
    \label{fig:mixture_models}
\end{figure*}

 
We design two mixture models. For the first model we create a normal distribution, $N$, such that its probability density blends smoothly into the log-normal distribution, $LN$, by using a mean value of $\mu_2=10$~s and standard deviation of $\sigma_2=2$\,s (\textit{model 1}). For the second model we design a normal distribution which has much less density overlap by using  a mean value of $\mu_2=15$~s and standard deviation of $\sigma_2=0.5$\,s (\textit{model~2}). For both models we compute the 25-year HDC as well as the 25-year IFORM contour. In model~1 the HDC and IFORM contour have similar shapes. Both have a concave path at high $T_z$-values and as expected the HDC is bigger in size (Fig. \ref{fig:mixture_models}\textit{a}). In contrast, model~2 has two distinct probability density maxima which lead to different shapes for the IFORM and HDC. While the HDC encloses two separated subregions the IFORM contour encloses a single contiguous region (Fig. \ref{fig:mixture_models}\textit{b}). This single region contains sea states with much lower probability densities than the conservative HDC as by its definition IFORM can only enclose one single contiguous region. Consequently, in this example an engineer who designs a structure to withstand all loads caused by the environmental states along this 25-year contour would design the structure to withstand some environmental states which are expected to occur extremely rarely. Therefore, possible structural designs which are limited by these environmental states would not be considered which could lead to bad design, either from a cost or engineering perspective.
\par 
\par
The apparent difference in shape between the two contours is interesting since it visually demonstrates that the IFORM contour does not have constant probability along its path and consequently does not enclose the most likely environmental states. Strictly, this should not be expected anyway, but since it is roughly true for many ordinary sea state models, one might intuitively interpret an IFORM contour that way. By IFORM's definition the  contour has two properties in the $U$-space: (i) constant probability density along its path and (ii) $\alpha$-probability halfspaces separated by lines which are tangent to the contour (Fig. \ref{fig:different_methods_calculation}a). Interestingly, for many sea state probability models these two properties roughly translate to the $X$-space. Here, we demonstrate the rough persistence of the constant probability density property for the unmodified sea state model since in this case IFORM and HDCs have similar shapes (Fig. \ref{fig:comparison_iform_huseby}\textit{b}). Rough persistence of the $\alpha$-halfspace property, on the other hand, has been shown by \citet{huseby:2013} who computed Monte Carlo contours which are defined by enforcing the $\alpha$-halfspace property in the original variable space (Fig. \ref{fig:different_methods_calculation}b). These Monte Carlo contours have been reported to have similar shapes as the IFORM contours. Thus, based on experience an engineer might intuitively interpret a typical IFORM contour to have roughly constant probability density and $\alpha$-halfspace exceedance probability in the original variable space.
\par 
This interpretation would not hold true for the multimodal model~2, however. In addition to clearly not having constant probability density it also does not roughly have $\alpha$-halfspace exceedance in the original variable space since the contour is concave. Not having any meaningful properties in the original variable space, raises the question how to intuitively interpret an IFORM contour in such a case. In contrast, the presented highest density contour with its constant probability density, $f_m$, along the contour and its enclosure of a probability of $1-\alpha$ offers a clear interpretation for any probability distribution. 
\section{Summary and conclusions}
In this work we present environmental contours which enclose regions of highest probability density. A highest density contour (HDC) has constant probability density along its path and occupies the smallest possible volume in the variable space for a given probability content. We compute the contour using numerical integration based on a grid, i.e. we iteratively find the minimum probability density, $f_m$, which leads to a contour containing the most likely environmental states which together have a probability of $1-\alpha$. Defined this way a $T$-year environmental contour is exceeded on average every $T$ years anywhere along the contour. This means precisely that such an environmental state is realized anywhere outside the environmental contour (and not in a further limited exceedance region). Highest density contours can be computed based on any probability density function, e.g. standard parametric sea state models, nonparametric models or extreme value models. The method's clear definition in terms of exceedance probability, $\alpha$, as well as its straightforward computation makes it an attractive alternative to the established IFORM approach.

\section*{Acknowledgements}
We thank R. Reisenhofer for fruitful discussions. This research did not receive any specific grant from funding agencies in the public, commercial, or not-for-profit sectors.

\bibliographystyle{elsarticle-harv}

\bibliography{highestdensitycontour}

\appendix
\section{Equations for \textit{p} dimensions}
We start by discretizing the $p$-dimensional probability space into $\Pi_{j=1}^p K_j$ grid cells with grid cell lengths of  $\Delta x_j$. Next, we calculate the cell-averaged probability density in each dimension, $\bar{f}_{Xj}$. This is done based on the respective cumulative distribution function, $F_{Xj}$:
\begin{equation}
       \bar{f}_{Xj}(x_j)=\dfrac{F_{Xj}(x_{j} + 0.5 \Delta x_j)-F_{Xj}(x_j -0.5 \Delta x_j)}{\Delta x_j}.
\end{equation}
Multiplying the $p$ individual probability densities yields the cell-averaged joint probability density, $\bar{f}$:
\begin{equation}
       \bar{f}(x_1, x_2, ..., x_p)=\Pi_{j=1}^p\bar{f}_{Xj}(x_j).
\end{equation}
Next, we compute the probability enclosed by a contour of $f_m$ probability density. This is done by calculating the sum of each cell's probability whose probability density is greater than or equal $f_m$: 
\begin{equation}
\begin{split}
    &\bar{F}(f_m)=\\
    &\sum_{k_1=1}^{K_1} \sum_{k_2=1}^{K_2} ... \sum_{k_p=1}^{K_p}
    \left\{
        \begin{array}{ll}
          \bar{f}(x_{1k_1}, x_{2k_2}, ..., x_{pk_p}) \Pi_{j=1}^p \Delta x_j &  \bar{f} \geq f_m\\
          0 & \bar{f} < f_m .
        \end{array}
    \right.
\end{split}
\end{equation}
Now, we can proceed as in two dimensions. We want to find the minimum probability density, $f_m$, that corresponds to the exceedance probability, $\alpha$, of interest:
\begin{equation}
\bar{F}(f_m)=1-\alpha
\end{equation}
As in two dimensions, this equation represents a root finding problem of a monotonically decreasing function ($\bar{F}(f_m)-1+\alpha=0)$ which can be solved with standard numerical methods, e.g. by using Matlab's \textit{fzero} function.
\par
\begin{figure*}
    \centering
    \includegraphics{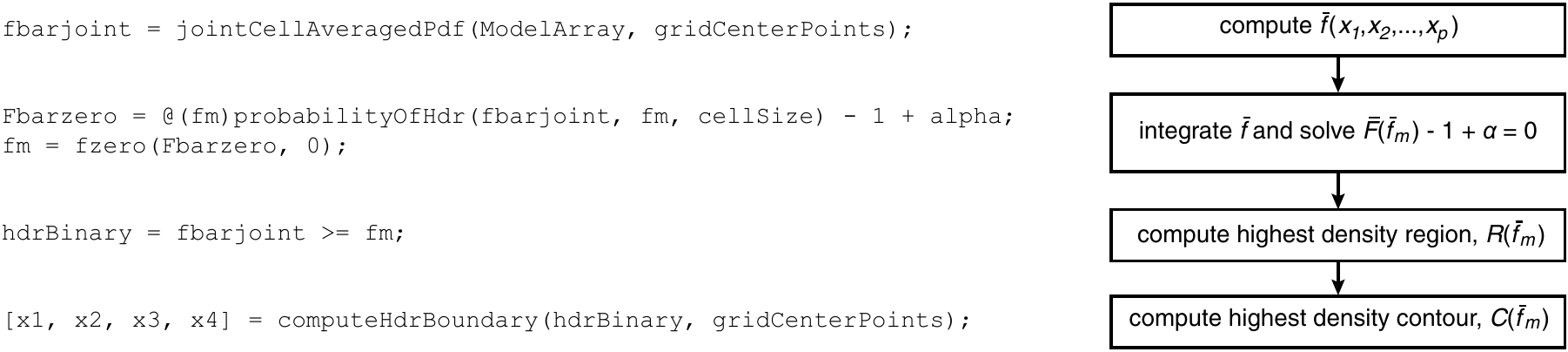}
    \caption{\textbf{Computer program to derive a highest density contour.} Left: Code snippet written in the Matlab programming language. Right: Corresponding flowchart.}
    \label{fig:code}
\end{figure*}

A Matlab implementation working up to four dimensions can be downloaded at \url{http://mathworks.com/matlabcentral/fileexchange/60876}. Figure \ref{fig:code} shows a source code snippet and the corresponding flowchart.
\par 
\end{document}